\documentclass[print]{ati}
\usepackage{graphicx,times}
\usepackage{array}
\usepackage{lmodern}
\usepackage[sort&compress, numbers,super]{natbib}
\bibpunct[,]{[}{]}{,}{n}{}{,}
\usepackage{url}
\usepackage[colorlinks=true,breaklinks=true, linkcolor=red,urlcolor=magenta,citecolor=blue,anchorcolor=green]{hyperref}
\usepackage{caption}
\usepackage{subcaption}
\usepackage{rotating}
\usepackage{CJKutf8}
\usepackage{amsmath,amssymb}
\usepackage{multirow}
\usepackage{threeparttable}
\renewcommand{\citet}[1]{\citeauthor{#1}(\citeyear{#1})\cite{#1}}

\newcommand{\eqn}[1]{Equation~(\ref{#1})}
\newcommand{\SecRef}[1]{Section~\ref{#1}}
\newcommand{\FigRef}[1]{Figure~\ref{#1}}
\newcommand{\TabRef}[1]{Table~\ref{#1}}

\newcommand{\ie}{\textit{i}.\textit{e}.}

\begin{document}

   \title{
   A Quasi-Optimal Stacking Method for Up-the-Ramp Readout Images
}
    
    \author{Guanghuan Wang
      \inst{1,2}
   \and Hu Zhan\correspondingAuthor{}
      \inst{1,3}
  \and Zun Luo
      \inst{1,2}
  \and Chengqi Liu
      \inst{3}
  \and Youhua Xu
      \inst{1}
   \and Chun Lin
       \inst{4}
   \and Yanfeng Wei
       \inst{4}
   \and WenLong Fan
       \inst{5}
   }
\correspondent{Hu Zhan}
\correspondentEmail{zhanhu@nao.cas.cn}
    \institute{Key Laboratory of Space Astronomy and Technology, National Astronomical Observatories, Chinese Academy of Sciences, Beijing 100101, China\\
          \and
              University of Chinese Academy of Sciences, Beijing 101408, China\\
          \and
              Kavli Institute for Astronomy and Astrophysics, Peking University, Beijing 100871, China\\
          \and
              National Key Laboratory of Infrared Detection Technologies, Shanghai Institute of Technical Physics, Chinese Academy of Sciences, Shanghai 200083,  China\\
          \and
              Shanghai Institute of Technical Physics, Chinese Academy of Sciences, Shanghai 200083, China\\
   }
   \date{Received:~March 26, 2023;   Accepted:~October 26, 2023;  Published Online:~December 26, 2023; 
            \DOI{xxxx-xxxx.0000-00} }
   \copyrights {2024~The Authors. }
   \citeinfo {Wang G. et al. 2024.}
   \abstract{
The non-destructive readout mode of a detector allows its pixels to be read multiple times during integration, generating a series of ``up-the-ramp'' images that keep accumulating photons between successive frames.
Since the noise is correlated across these images, an optimal stacking generally requires weighting them unequally to achieve the best signal-to-noise ratio (SNR) for the target.
Objects in the sky show wildly different brightness, and the counts in the pixels of the same object also span a wide range. 
Therefore, a single set of weights cannot be optimal for all cases. 
To keep the stacked image more easily calibratable, however, we choose to apply the same weight to all the pixels in the same frame. 
In practice, we find that the results of high-SNR cases degrade only slightly by adopting weights derived for low-SNR cases, whereas the low-SNR cases are more sensitive to the weights applied.
We therefore propose a quasi-optimal stacking method that maximizes the stacked SNR for the case of SNR=1 per pixel in the last frame and demonstrate with simulated data that it always enhances the SNR more than the equal-weight stacking method and the ramp fitting method. 
Furthermore, we give an estimate of the improvement of limiting magnitudes for the China Space Station Telescope (CSST) based on this method. 
Compared with the conventional readout mode, which is equivalent to taking the last frame of the non-destructive readout, stacking $30$ up-the-ramp images can improve the limiting magnitude by about $0.5~\mathrm{mag}$ for CSST near-infrared observations, effectively reducing the readout noise by about $62\%$. 
  \keywords{Astronomical Astronomical detectors(84) --- Infrared observatories(791) --- Astronomy data reduction(1861) --- Astronomy image processing(2306)}}

   \authorrunning{ASTRONOMICAL TECHNIQUES \& INSTRUMENTS }

   \titlerunning{Wang et al.: quasi-optimal Stacking for Ramp Images}
   \VolumeNumberPageYear{2}{6}{1}{2024}
   \MonthIssue{Jan}
   \DOItail{20230915.006}
   \maketitle
   \setcounter{page}{\Page}

\section{Introduction}
\label{sect1}

Infrared detectors are often designed with a non-destructive readout mode, which can sample the signal in a pixel multiple times during integration without altering it. 
The non-destructive readout mode is  extensively utilized in astronomical infrared cameras, such as the Near-Infrared Camera and Multi-Object Spectrometer and Wide Field Camera 3 (WFC3) on the Hubble Space Telescope, the Infrared Array Camera of the Spitzer Space Telescope, all infrared cameras of the James Webb Space Telescope, the Near Infrared Spectrograph and Photometer of the Euclid mission, and the Wide Field Instrument on board the forthcoming Nancy Grace Roman Space Telescope \cite{skinner1998orbit, baggett2008wide, fazio1998infrared, rieke2007infrared, corcione2012board, munoz2021euclid, casertano2022determining, rauscher2019principal}. 
The China Space Station Telescope \citep[CSST, also known as the Xuntian Space Telescope;][]{Zhan2011,Zhan2021,Gong2019} will also observe in the near infrared (NIR) with detectors made by Shanghai Institute of Technical Physics \cite{LiangQinghua2024}. 

The non-destructive readout mode has several advantages. First of all, multiple frames that are sampled during integration can be reduced to a single frame to mitigate the impact of the readout noise and improve the data quality \cite{sullivan2014near}. On top of that, the data loss caused by cosmic rays can be at least partially recovered in the non-destructive readout data. Pixels struck by cosmic rays exhibit jumps in their integration ramps, and proper algorithms can identify and remove these jumps \cite{anderson2011optimal, offenberg2001validation, 2014SPIE.9143E..3ZR, fixsen2000cosmic}. Lastly, the pixel values read out before saturation can be used for non-linearity correction to increase the detector's effective dynamic range, enabling richer details to be captured \cite{darson2017real, Pengjunhua2010NDR}. 

The sequence of output frames with increasing exposure time in non-destructive readout mode forms a three-dimensional data cube. Hereafter, we refer to this series of images as up-the-ramp images, or ramp images for short. A reduction strategy is needed to combine them into a single image to enhance the signal-to-noise ratio (SNR). 
An equal-weight stacking method is adopted by WFC3 to reduce the effective readout noise \cite{2023wfci.book...15D}. Since the SNR increases with the exposure time, weighting all the frames equally biases against high-SNR ones, causing the stacked SNR even below the SNR of the last frame in the signal-dominant regime. 
One may also fit the integration ramp before saturation for each pixel and obtain the value at the exposure time of the last frame to construct a ``combined'' image  \cite{gordon2005reduction, garnett1993multiply, robberto2007analysis}, though this ramp fitting method is unstable with insufficient number of up-the-ramp exposures. 
In this paper, we develop a weighted stacking method that maximizes the SNR of the stacked image for a low-SNR target case while maintaining good performance over the whole SNR range. 
Simulations demonstrate that our quasi-optimal stacking method achieves higher SNR than the equal-weight stacking method and the ramp fitting method do and can be particularly helpful for faint object detection with observations in the readout-noise dominant regime.

The rest of the paper is organized as follows. In \SecRef{sect2}, we derive the optimal weights and examine their performance when the target case mismatches actual observations. We then compare our method with the commonly used reduction methods under different conditions in \SecRef{sect3}. \SecRef{sect4} provides an estimate of improvement by the quasi-optimal stacking method for CSST NIR observations. The conclusions are drawn in \SecRef{sect5}. 

\section{Quasi-optimal Stacking}
\label{sect2}

In a series of $N$ ramp images, the pixel values in each frame are accumulated on top of the previous one. 
Although the optimal stacking weights depend on the signal and noise, which vary from pixel to pixel, 
we choose to apply the same weight to all the pixels in a single frame. With this approach, the stacking is not optimal for all pixels, but flux calibration of the stacked image can be done in the usual way. 
If the weights are optimized for each pixel individually according to the signal and noise in it, which are not known a priori in general, stacking results in low SNR regions can be significantly affected by random fluctuations of the noise and potentially bias flux calibration.

We start with the case of a single pixel for simplicity.One can also think of this case as a ramp series of flat-field images with identical pixels.
Assuming the bias correction and non-linear correction have been done already, we have the pixel value $f_i$ in the $i$-th frame
\begin{equation}\label{eq_fi}
    f_i =  s_i + \Delta s_i + b_i + \Delta b_i + \Delta r_i,
\end{equation}
where $s_i$ and $\Delta s_i$ are the signal of the source and associated Poisson fluctuation, $b_i$ and $\Delta b_i$ are the background (including contributions from dark current and the sky background) and associated Poisson fluctuation, and $\Delta r_i$ is the Gaussian fluctuation due to the readout noise. 
The circuit gain is irrelevant in the stacking process and is omitted from \eqn{eq_fi} without losing generality. 
The signal $s_i$ and the background $b_i$ increase linearly with time.
The fluctuation terms, $\Delta s_i$, $\Delta b_i$ and $\Delta r_i$, are all independent of each other, so that the covariance of the pixel values between different frames becomes
\begin{eqnarray}
    \nonumber
    \bm{C} &\equiv& \mathrm{Cov}(f_i, f_j)  \\ \nonumber
    &=& \mathrm{Cov}(\Delta s_i, \Delta s_j) + \mathrm{Cov}(\Delta b_i, \Delta b_j) +\mathrm{Cov}(\Delta r_i, \Delta r_j) \\
    \nonumber
    &=& s_{\mathrm{min}(i,j)}+b_{\mathrm{min}(i,j)}+r^2 \delta_{ij},
\end{eqnarray}
where $\delta_{ij}$ is the Kronecker Delta function, and the readout noise $r$ is assumed to be the same in all frames.
In matrix form, we have 
\begin{equation}\label{eq_c}\nonumber
\bm{C}  = \left[ {
    \begin{array}{cccc}
        s_1+b_1+r^2 &s_1+b_1    &\cdots &s_1+b_1\\
        s_1+b_1     &s_2+b_2+r^2&\cdots &s_2+b_2\\
        \vdots      &\vdots     &\ddots &\vdots \\
        s_1+b_1     &s_2+b_2    &\cdots &s_N+b_N+r^2\\
    \end{array} } \right].
\end{equation}
The readout noise is not correlated between frames and appears only on the diagonal of the covariance matrix. The values of the off-diagonal elements correspond to the Poisson noise in the previous frame, which is not correlated with the signal or the background accumulated in later frames.

We apply a series of weights $\bm{\omega}=[\omega_1,\omega_2,\cdots,\omega_N]^\mathrm{T}$ to the ramp images and obtain the stacked pixel value $\bm{\omega}^\mathrm{T} \bm{f}$, where $\bm{f}=[f_1, f_2, \cdots, f_N]^\mathrm{T}$.
The SNR of the stacked pixel can be expressed as
\begin{equation}\label{eq_snrws}
\mathrm{SNR} = \frac{\bm{\omega}^\mathrm{T} \bm{s}}
           {\sqrt{\bm{\omega}^\mathrm{T}\bm{C}\bm{\omega}}},
\end{equation}
where $\bm{s}=[s_1,s_2,\cdots,s_N]^\mathrm{T}$. 
To find the weights that maximize the SNR, we set the derivative of the SNR with respect to the weights to zero, \ie{},
\begin{equation}\label{eq_derivation}
\frac{\partial\mathrm{SNR}}{\partial\bm{\omega}} = 
    \frac{\bm{s}}{\sqrt{\bm{\omega}^\mathrm{T}\bm{C}\bm{\omega}}}
    - \frac{\bm{\omega}^\mathrm{T} \bm{s}}
        {(\sqrt{\bm{\omega}^\mathrm{T}\bm{C}\bm{\omega}})^3}\ \bm{C}\bm{\omega}=0. 
\end{equation}
Since both $\bm{\omega}^\mathrm{T}\bm{C}\bm{\omega}$ and $\bm{\omega}^\mathrm{T}\bm{s}$ are scalars, the optimal weight vector must satisfy
\begin{equation}
    \bm{s} \propto \bm{C}\bm{\omega}_\mathrm{opt}.
\end{equation}
The solution is therefore

\begin{equation}\label{eq_wopt}
    \bm{\omega}_\mathrm{opt} \propto \bm{C}^{-1}\bm{s}
\end{equation}
with the maximum SNR
\begin{equation}
\mathrm{SNR}_\mathrm{opt} = \sqrt{\left(\bm{C}^{-1}\bm{s}\right)^\mathrm{T}\bm{s}}.
\end{equation}
The weights are determined up to a scaling factor in \eqn{eq_wopt}. A proper normalization of  $\bm{\omega}_\mathrm{opt}$ is to require the ``stacked exposure time'' to match the exposure time of the last frame, i.e.,
\begin{equation}\label{eq_wnorm}
\bm{\omega}_\mathrm{opt}^\mathrm{T}\bm{t}=t_N,
\end{equation}
where $\bm{t}=[t_1,t_2,\cdots,t_N]^\mathrm{T}$ contains the exposure time of each frame. 
The count rate of the signal is then $S=s_i/t_i$, and that of the background is $B=b_i/t_i$. 
We assume hereafter for convenience that the frames are equally spaced in time, \ie, $t_i=i\times t_1$.

\begin{table*}[ht]
    \begin{minipage}[t]{1.0\linewidth}
    \caption[ ]{Description of weights and SNRs in this paper}
    \label{tab0}
    \end{minipage}
    \centering
    \renewcommand{\arraystretch}{1.5}
    \begin{threeparttable}
        \begin{tabular}{l p{0.8\textwidth}}
        \hline\noalign{\smallskip}
        \multicolumn{1}{c}{Symbol} & \multicolumn{1}{c}{Description} \\
        \hline\noalign{\smallskip}
        $\bm{\omega}_\mathrm{opt}$      & The optimal weight vector that maximizes the SNR per pixel in the stacked image \\
        \noalign{\smallskip}
        $\bm{\omega}_\mathrm{target}$   & The optimal weight vector derived for a ramp series in which the SNR per pixel in the last frame equals $\mathrm{SNR}_\mathrm{target}$ \\
        \noalign{\smallskip}
        $\mathrm{SNR}_\mathrm{last}$    & The SNR per pixel in the last frame\\
        \noalign{\smallskip}
        $\mathrm{SNR}_\mathrm{target}$  & The SNR per pixel in the last frame that is used as a target to derive the optimal weight vector  $\bm{\omega}_\mathrm{target}$\\
        \noalign{\smallskip}
        $\mathrm{SNR}_{\mathrm{opt}}$   & The optimal SNR achieved when $\mathrm{SNR}_\mathrm{target}=\mathrm{SNR}_\mathrm{last}$  \\
        \noalign{\smallskip}
        $\mathrm{SNR}_\mathrm{stack}(\bm{\omega}_\mathrm{target})$ & The SNR per pixel in the image stacked with $\bm{\omega}_\mathrm{target}$\\
        \hline
        \end{tabular}
    \end{threeparttable}
\end{table*}

\begin{figure}
    \begin{minipage}[t]{0.9\linewidth}
    \centering
    \includegraphics[width=1.0\textwidth]{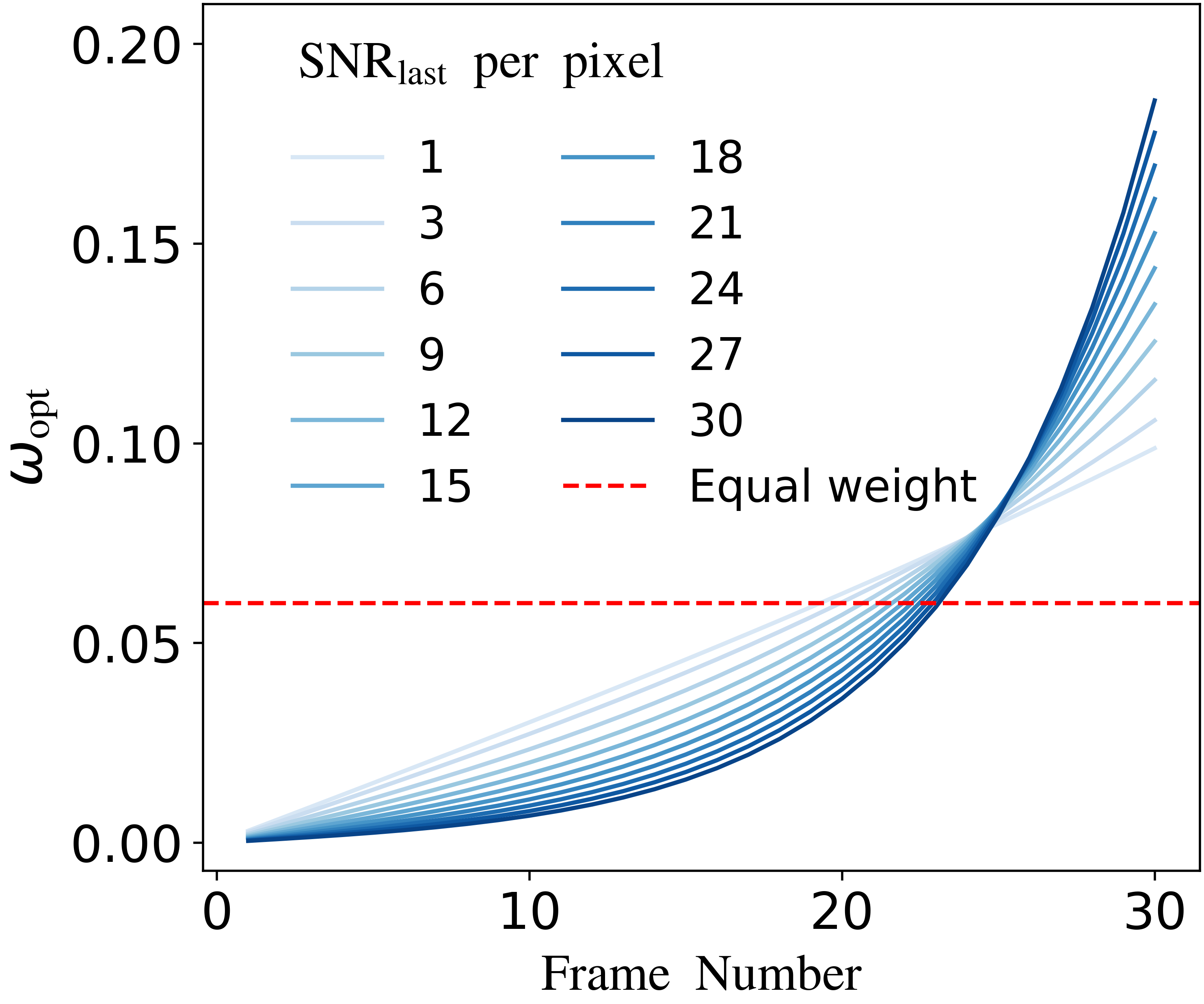}
    \caption{Optimal stacking weights. Lines of different shades represent different signal levels in the last frame. The green dashed line represents equal-weighted stacking. 
    Each set of weights is normalized according to \eqn{eq_wnorm}. 
    }
    \label{Fig1}
    \end{minipage}
\end{figure}

\FigRef{Fig1} displays the optimal weights for a series of ramp images with the same setting ($N=30$, $b_i=0$ and $r=50~\mathrm{e^-}$) except the signal level.  
The frames are stacked in the same way as the stacking of a single pixel. 
Each line in the figure represents a different case of SNR per pixel in the last frame, $\mathrm{SNR_{last}}$, defined by
\begin{equation}\label{eq_sl}
\mathrm{SNR_{last}}=\frac{s_N}{\sqrt{s_N+b_N+r^2}}. 
\end{equation}
The figure shows that, as $\mathrm{SNR_{last}}$ increases, more weight is assigned to later frames. 

\begin{figure}
    \begin{minipage}[t]{0.95\linewidth}
    \centering
    \includegraphics[width=1.0\textwidth]{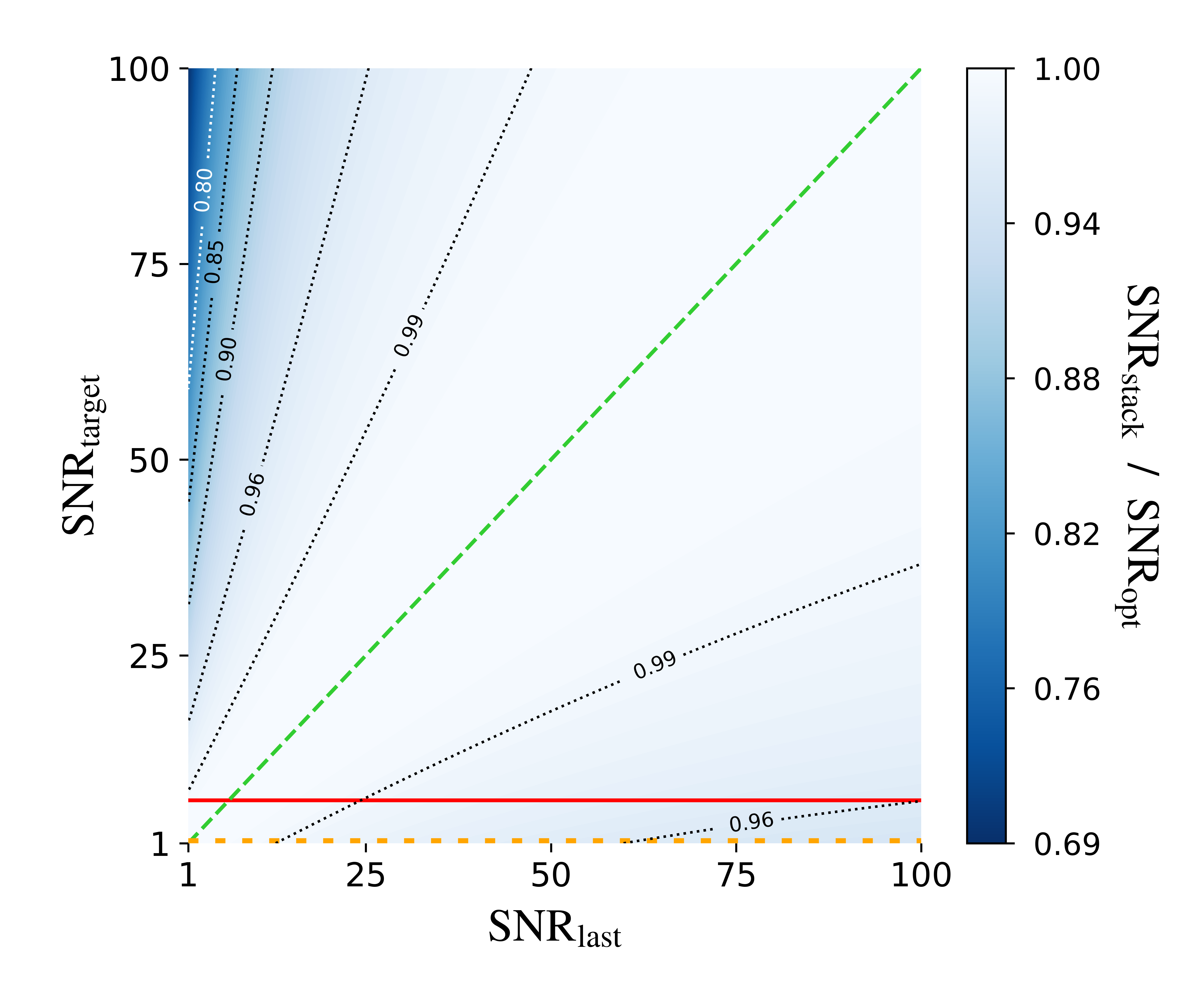}
    \caption{SNR degradation of stacking with non-optimal weights. The green dashed line represents the optimal case in which $\mathrm{SNR_{target}}$ always matches $\mathrm{SNR_{last}}$. The red solid line marks $\mathrm{SNR_{target}}=6.5$. 
    The orange dotted line marks $\mathrm{SNR_{target}}=1$. 
    }
    \label{Fig2}
    \end{minipage}
\end{figure}

In real observations, counts in pixels can be wildly different, so that one set of weights cannot maximize the SNR for all. 
However, to keep the finale image more easily calibratable, all pixels in the ramp images should be stacked with the same set of weights. Therefore, one can only choose a particular case of $\mathrm{SNR_{last}}$ as the target, which we label as $\mathrm{SNR_{target}}$, and use the optimal weights $\bm{\omega}_\mathrm{target}$ derived for the target to stack the ramp images.
To explore the effect of target setting, we expand the $\mathrm{SNR_{last}}$ range of the examples in Figure~\ref{Fig1} to $1 \le \mathrm{SNR_{last}} \le 100$ and stack the ramp images with $\bm{\omega}_\mathrm{target}$. 
\eqn{eq_snrws} is then used to calculate the SNR of the stack, $\mathrm{SNR_{stack}}(\bm{\omega}_\mathrm{target})$. 
The result is normalized by the matched case of $\mathrm{SNR_{target}}=\mathrm{SNR_{last}}$ and is shown as a function of $\mathrm{SNR_{last}}$ and $\mathrm{SNR_{target}}$ in \FigRef{Fig2}
(see \TabRef{tab0} for a description of symbols related to various weights and SNRs).

As indicated by the values in the top-left corner of \FigRef{Fig2}, setting a brighter target case (higher $\mathrm{SNR_{target}}$) for fainter objects (lower $\mathrm{SNR_{last}}$) results in a significant decrease in the SNR of the stacked frame. A strategy of setting a fainter target case can avoid falling into this circumstance and can achieve an SNR almost the same as that using the true optimal weights $\bm{\omega}_\mathrm{opt}$ (as marked by the green dashed line). With $\mathrm{SNR_{target}}$ set to 6.5 (red solid line), the decrease of SNR compared to using the true optimal weights is less than $4\%$ for $\mathrm{SNR_{last}} \lesssim 100$, and with $\mathrm{SNR_{target}}=1$ (horizontal axis), the degradation does not exceed $1\%$ for $\mathrm{SNR_{last}}\lesssim 10$. For both cases, the resulting $\mathrm{SNR_{stack}}$ is still higher than $\mathrm{SNR_{last}}$ (see similar cases in \SecRef{sect3}).

\begin{table*}[!ht]
    \begin{minipage}[t]{0.999\linewidth}
    \caption[ ]{Parameters for generating flat-field ramp images}
    \label{tab1}
    \end{minipage}
    \begin{center}
    \begin{threeparttable}
    \begin{tabular}{l c c c c}
        \hline\noalign{\smallskip}
        $\rm{Dominant~noise}$ & $r$ & $B$ & $S$ & $\mathrm{SNR_{last}^{~1}}$\\
        \hline\noalign{\smallskip}
        $\rm{Readout}$  & $50~\mathrm{e^-}$ & $~6~\mathrm{e^-/s}$ & $0.39$--$4~\mathrm{e^-/s}$ & $1.0$--$9.5$\\
        \noalign{\smallskip}
        $\rm{Background}$     & $25~\mathrm{e^-}$ & $12~\mathrm{e^-/s}$& $0.33$--$4~\mathrm{e^-/s}$ & $~1.0$--$10.9$\\
        \noalign{\smallskip}
        $\rm{Source}$& $50~\mathrm{e^-}$ & $~6~\mathrm{e^-/s}$ &$23$--$100~\mathrm{e^-/s}$  & $~41.7$--$110.6$\\
        \hline
    \end{tabular}
    \begin{tablenotes}
        \footnotesize
        \item[1] The accumulated exposure time of the last frame is $t_N=150$~s.
    \end{tablenotes}
    \end{threeparttable}
    \end{center}
\end{table*}

Given that the detection limit is crucial to most projects, one can boost the SNR of the faintest objects detectable in the final stack by setting a low target $\mathrm{SNR_{target}}$.  
This approach is not optimal for all SNR cases, so we refer to it as the quasi-optimal stacking method. For a telescope whose point spread function (PSF) is properly sampled, the most compact objects like stars usually cover more than 10~pixels each. If one adopts SNR$\ge5$ as the threshold, then the average SNR per pixel would be less than 2 for a barely detected star. 
With extended sources like galaxies, the same threshold could mean an average SNR per pixel approaching unity or even less.
Since the $\mathrm{SNR_{stack}}$ of the stacked image is insensitive to the target $\mathrm{SNR_{target}}$ for low SNR cases, a reasonably good strategy would be setting the target $\mathrm{SNR_{target}} = 1$ for optimization.

\section{Comparison of Reduction Methods}
\label{sect3}

As mentioned in \SecRef{sect1}, conventional methods for ramp-image reduction include the equal-weight stacking method and the ramp fitting method. 
The latter fits the slope of the integration ramp for each pixel, which represents the count rate in it, and the count rates of all pixels form the final image \cite{garnett1993multiply, rauscher2007detectors, robberto2007analysis}. 
The last frame of a ramp series is equivalent to a frame of the same total exposure time taken in the usual destructive readout mode. It serves as a useful benchmark for comparison of different ramp-image reduction methods, and we also refer to it as a reduction method for convenience.
To evaluate the performance of the quasi-optimal stacking method and the other methods, we carry out two tests with simulated images: the flat-field test and the point-source test. 
Details are given below. 

\subsection{Flat-field test}
\label{sec31}

For the flat-field test, we generate ramp images with uniform illumination by setting the signal $s_i=t_i S$ to the same value for all pixels in the same frame. A uniform background $b_i=t_i B$ is similarly applied. The readout noise $r$ is assumed to be the same for all pixels and all frames. 
The images have a size of $2000\times2000$~pixels each, large enough to keep statistical errors negligible. Randomly drawn values from Poisson distribution of the signal, that of the background and Gaussian distribution of the readout noise are then summed to give the count in each pixel. 

\eqn{eq_fi} shows that there are three types of noise in the pixel: Poisson (or photon) noise of the source, Poisson noise of the background and Gaussian noise of the readout process. 
Ramp-image reduction methods can behave differently in regimes dominated by different noise, so we design the test accordingly in three cases with parameters listed in Table \ref{tab1}. 
To be specific, by dominant noise we mean that its variance is greater than that of the other two types of noise combined.

\begin{figure*}[!htb]
    \includegraphics[width=1.0\textwidth,angle=0,scale=1.0]{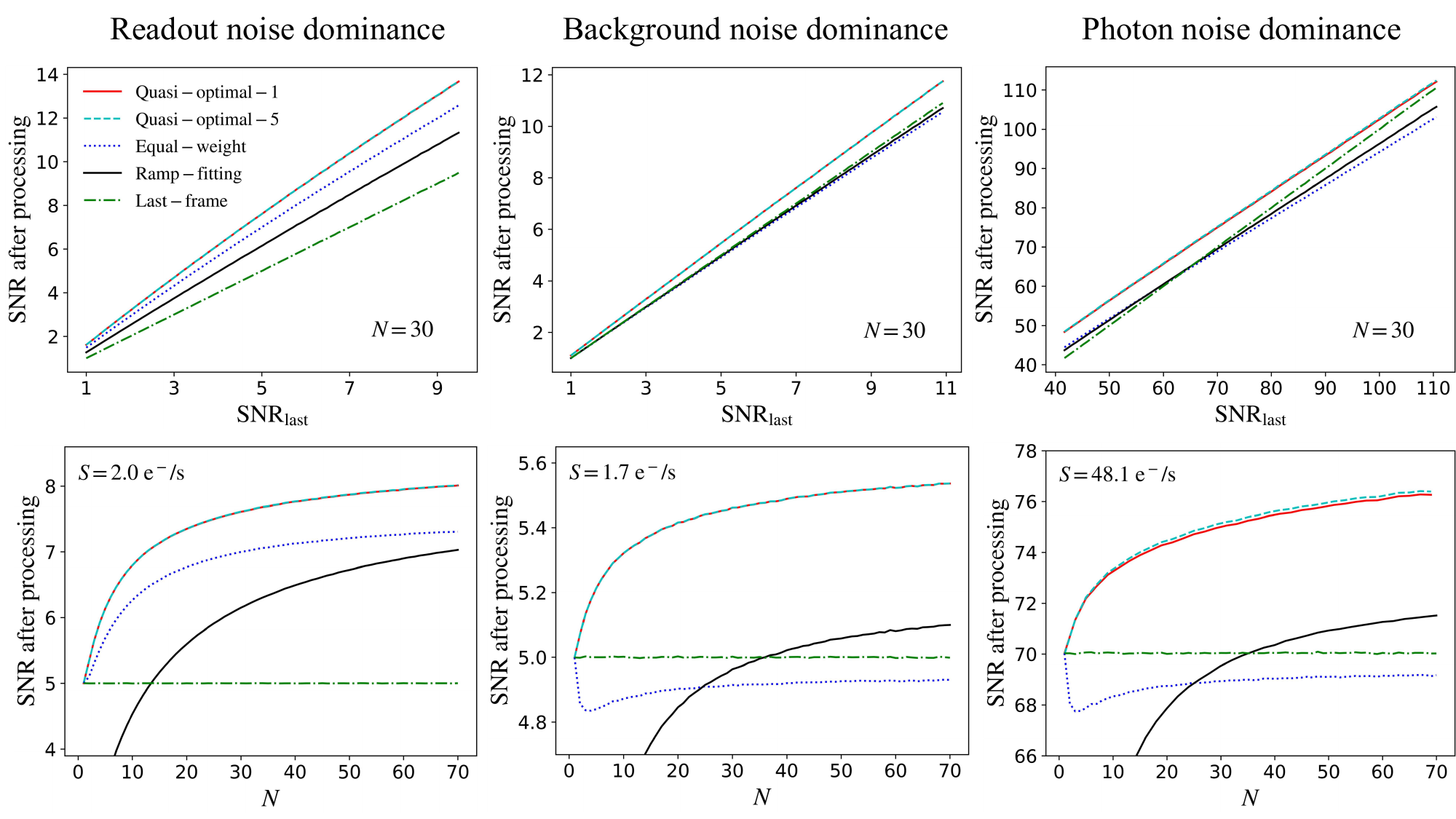}
\begin{minipage}[]{1.0\textwidth}
    \caption[]{SNR per pixel of reduced flat-field images with the quasi-optimal stacking method (red solid curve for $\mathrm{SNR_{target}}=1$ and green dashed curve for $\mathrm{SNR_{target}}=5$), the equal-weight stacking method (blue dotted curve), the ramp fitting method (black solid curve) and the last-frame method (green dash-dotted curve).
    The columns correspond to different cases of dominant noise, as indicated by the text at the top of the image. The first row compares the SNRs of the reduced images as the count rate of the signal source $S$ and the resulting  $\mathrm{SNR_{last}}$ increase, while the second row illustrates the dependence of the SNRs on the number of frames $N$ in the ramp series with $t_N$ fixed. 
    }
    \label{Fig3}
\end{minipage}
\end{figure*}

The first row of \FigRef{Fig3} compares the SNR per pixel achieved by the four ramp-image reduction methods.
The SNR is given by the ratio of the standard deviation of all pixels in the reduced image to the mean of these pixels.
It is seen that our quasi-optimal stacking method consistently outperforms the other methods in all the three cases. 
Since the quasi-optimal stacking method assigns more weight to frames with more signal, it should always obtain a better SNR than the equal-weight stacking method does. 
Moreover, the results of $\mathrm{SNR_{target}}=1$ (red solid curve) and $\mathrm{SNR_{target}}=5$ (green dashed curve) are essentially identical, supporting the strategy of setting 
$\mathrm{SNR_{target}}=1$.
The last-frame method actually surpasses the quasi-optimal stacking method at $\mathrm{SNR_{last}}\gtrsim 130$, which is beyond the range shown in the photon-noise dominant panel.
A small degradation to such a high SNR can hardly affect object detection, and, hence, setting 
$\mathrm{SNR_{target}}$ to unity can be considered a fail-safe choice for the quasi-optimal stacking method. 

Although one usually tries not to observe in the readout-noise dominant regime, it is sometimes unavoidable. Actually, in this regime, we see more enhancement of the SNR over the last-frame method by the other three methods than that in the background-noise dominant regime and the photon-noise dominant regime (collectively, the Poisson-noise dominant regime). 
This indeed manifests as one of the advantages of the non-destructive readout mode.

To check the dependence of the SNR of the reduced image on the number of non-destructive readouts during the entire exposure time, we generate another set of ramp images according to \TabRef{tab1} but with $N$ ranging from 1 to 70 and $S$ fixed.
The results are shown in the second row of \FigRef{Fig3}. The stacking methods and the ramp fitting method show convergent behavior as $N$ becomes large enough.
The quasi-optimal stacking method achieves the best SNR among all the methods and also robustly outperforms the last-frame method in all the three cases shown.
The equal-weight stacking method is slightly worse than the last-frame method in the Poisson-noise dominated regime. The ramp fitting method performs poorly if the number of frames is not large enough.
This is consistent with the noise of the ramp fitting method $\sigma_\mathrm{RF}$ derived in Ref~\cite{robberto2007analysis}:
\begin{equation} \label{sig_rf_rnd}
    \sigma^2_\mathrm{RF}=\frac{12N}{N^2-1}r^2
\end{equation}
in the readout-noise dominated regime and
\begin{equation} \label{sig_rf_pnd}
    \sigma^2_\mathrm{RF}=\frac{6}{5}\frac{N^2+1}{N^2-1}s_N
\end{equation}
in the Poisson-noise dominated regime. 
With a large number frames, the noise of the ramp fitting method approaches a factor of $\sqrt{12}$ larger than that of the equal-weight stacking method in the readout-noise dominated regime according to \eqn{sig_rf_rnd} and is slightly higher than that of the last frame in the Poisson-noise dominated regime according to \eqn{sig_rf_pnd}.

\begin{figure*}[!htb]
    \centering
    \includegraphics[width=1.0\textwidth,angle=0,scale=0.9]{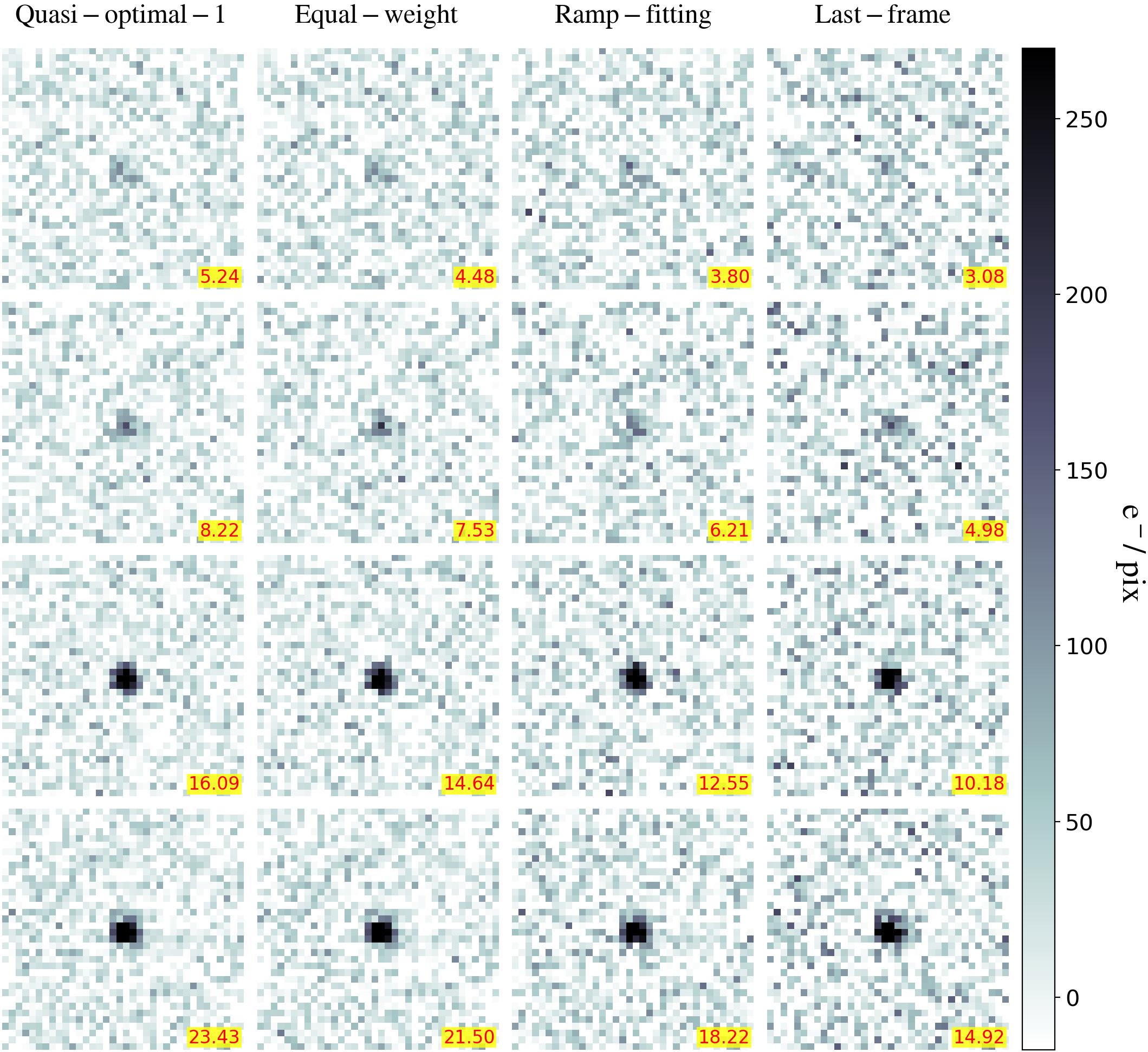}
\begin{minipage}[]{1.0\textwidth}
    \vspace{3mm}
    \caption[]{Mock stellar images reduced from the ramp series using different methods. From left to right, the panels in each row present the results of the quasi-optimal stacking method ($\mathrm{SNR_{target}}=1$), the equal-weight stacking method, the ramp fitting method and the last-frame method. The SNR of the star in an aperture of 2.3~pixels (encircling $\sim 80\%$ energy) is labeled in each panel. From top to bottom, the total electron counts of the star in each row are set to $891~\mathrm{e^-}$, $1492~\mathrm{e^-}$, $3016~\mathrm{e^-}$ and $4572~\mathrm{e^-}$, corresponding to simulation input SNRs of 3, 5, 10 and 15, respectively, in the last frame.
    }
    \label{Fig4}
\end{minipage}
\end{figure*}

\subsection{Point-source test}
\label{sec32}

The flat-field test in the previous subsection is fairly instructive, but it does not encompass the complexity of sky images. 
Here we provide a test based on ramp images of a circular Gaussian PSF at different SNRs, which is designed to mimic real observations of point-sources like stars. 
The full width at half maximum (FWHM) of the PSF is set to $3$ pixels, and the background count rate $B$, the readout noise $r$ and the total exposure time $t_N$ assume the same values of the readout-noise dominant case in \TabRef{tab1}.
The ramp images consists of $N=30$ frames and are processed with the quasi-optimal stacking method ($\mathrm{SNR_{target}}=1$), the equal-weight stacking method, the ramp fitting method and the last-frame method. 
We then make aperture photometry measurements on the reduced images using the Python Library for Source Extraction and Photometry (\texttt{SEP}) \cite{barbary2016sep, bertin1996sextractor}. 

The reduced images are shown in \FigRef{Fig4}. The star is at the center of each panel. Its signal is effectively all contained within 5 pixels from the center, so the region outside the star's footprint is essentially a map of the background and the readout noise. 
The number labeled in each panel is the star's SNR within an aperture of 2.3~pixels. It is seen that the quasi-optimal stacking method can boost the star's SNR by $57\sim 70\%$ over that in the last frame. Such an enhancement can make a difference between detection and non-detection. 
For example, in the first row of \FigRef{Fig4}, the star is hardly distinguishable from the background in the last-frame image (the rightmost panel) but becomes more discernible in the images processed by the other three methods, in the same rank of performance as seen in the readout-noise dominant case in \FigRef{Fig3}.
The visual impression is consistent with the SNR labeled in each panel, which demonstrates quantitatively that the quasi-optimal stacking method offers the best chance for detecting faint objects close to the detection threshold. 
It is also worth noting that the enhancement of the SNR is in part due to the suppression of the readout noise, which can be visually confirmed for the two stacking methods.

\section{Improvement for the CSST}
\label{sect4}

\begin{table*}
    \caption[]{Specifications of the CSST and its NIR imager}
    \label{tab2}
    \begin{center}
    \begin{threeparttable}
    \begin{tabular}{|c|c|c|cc|}
    \hline
    Telescope aperture  & $2~\mathrm{m}$    & Pixel scale & \multicolumn{2}{c|}{$0.11''\times0.11''$}     \\ \hline
    System throughput & $\ge50.4\%$       & Filter bandpass       & \multicolumn{1}{c|}{$\mathrm{J'}$: $0.9$--$1.3~\mathrm{\mu m}$}  & $\mathrm{H'}$: $1.3$--$1.7~\mathrm{\mu m}$  \\ \hline
    Readout noise & $\le50~\mathrm{e^-}$& Sky background$^{~1}$    & \multicolumn{1}{c|}{$0.74~\mathrm{e^-/(pix\cdot s)}$} & $0.54~\mathrm{e^-/(pix\cdot s)}$ \\ \hline
    Dark current  & $\le5~\mathrm{e^-/s}$ & $R_{80}$ & \multicolumn{1}{c|}{$0.21''$} & $0.26''$ \\ \hline
    \end{tabular}
    \begin{tablenotes}
    \footnotesize
    \item[1] The sky background is calculated for the CSST using the spectral energy distribution from the instrument handbook of WFC3 \cite{2023wfci.book...15D}. 
    \end{tablenotes}
    \end{threeparttable}
    \end{center}
\end{table*}

\begin{table*}
    \caption[]{Improvement of the limiting magnitude and effective readout noise with the number of frames taken in the ramp series}
    \label{tab3}
    \centering
    \begin{tabular}{c cc cc}
    \hline
    \textbf{N}  & \multicolumn{2}{c}{\textbf{Limiting~magnitude}} & \multicolumn{2}{c}{\textbf{Effective~readout~noise}} \\ 
    \hline
    & \multicolumn{1}{c}{($\mathrm{J'}$ band)} & ($\mathrm{H'}$ band) & \multicolumn{1}{c}{($\mathrm{J'}$ band)} & ($\mathrm{H'}$ band) \\ 
    1   & \multicolumn{1}{c}{23.72} & 23.16 & \multicolumn{1}{c}{50.00} & 50.00 \\
    10  & \multicolumn{1}{c}{24.11} & 23.55 & \multicolumn{1}{c}{27.73} & 27.64 \\
    20  & \multicolumn{1}{c}{24.22} & 23.66 & \multicolumn{1}{c}{21.84} & 21.73 \\
    30  & \multicolumn{1}{c}{24.26} & 23.71 & \multicolumn{1}{c}{19.09} & 18.97 \\
    40  & \multicolumn{1}{c}{24.29} & 23.74 & \multicolumn{1}{c}{17.41} & 17.29 \\
    50  & \multicolumn{1}{c}{24.31} & 23.75 & \multicolumn{1}{c}{16.24} & 16.12 \\
    60  & \multicolumn{1}{c}{24.32} & 23.77 & \multicolumn{1}{c}{15.36} & 15.24 \\
    70  & \multicolumn{1}{c}{24.33} & 23.78 & \multicolumn{1}{c}{14.66} & 14.55 \\ \hline
    \end{tabular}
\end{table*}

In this section, we assess the improvement of CSST NIR imaging observations in terms of the limiting magnitude and effective readout noise with the quasi-optimal stacking method. 

The photoelectron count of a star of magnitude $m_\mathrm{AB}$ in the CSST NIR imager is given by
\begin{equation}\label{eq_nobj}
s_{80} = 0.8 A_\mathrm{apr}~ T_\mathrm{sys}~ t_N \int\frac{f_\nu(m_\mathrm{AB})}{h\nu}d\nu, \\
\end{equation}
where $s_{80}$ is the count within the radius encircling 80\% energy of the PSF ($R_{80}$), $A_\mathrm{apr}$ is the area of the telescope's aperture, $T_\mathrm{sys}$ is the system throughput (assumed to be constant),  $f_\nu(m_\mathrm{AB})$ is the spectral flux density of the star, and the integration limits are defined by the filter cut-on and cut-off frequencies.
By definition of the AB magnitude system, the spectral flux density $f_\nu$ is flat over frequencies and is related to $m_\mathrm{AB}$ by
\begin{equation}\label{eq_mab}
    m_\mathrm{AB} = -2.5\log_{10}\left( \frac{f_\nu}{f_0} \right),
\end{equation}
where $f_0 = 3.631\times10^{-23}~\mathrm{W~ m^{-2} Hz^{-1}}$. 
The exposure time $t_N$ is fixed to the nominal value of 150~s. 
The SNR of the star is determined from the reduced image of simulated ramp images in the same way as done in \SecRef{sec32} except that relevant parameters are adjusted to the values in \TabRef{tab2}. 
We then obtain the limiting magnitude corresponding to the detection threshold of SNR=5.

Table \ref{tab3} summarizes the improvement of the limiting magnitude and effective readout noise in $\mathrm{J'}$ and $\mathrm{H'}$ bands as the number of frames in the ramp series ($N$) increases. 
The CSST NIR imager is able to output a maximum of roughly 70 frames during each 150~s exposure, and hence \TabRef{tab3} stops at $N=70$. 
In both bands, stacking $30$ frames with the quasi-optimal stacking method can increase the limiting magnitude by about $0.5~\mathrm{mag}$ compared to the result of the last frame, which is the same as the $N=1$ case, and stacking $70$ frames can increase it by about $0.6~\mathrm{mag}$. 
Given that the limiting magnitude increases only slowly for $N\ge 30$, we recommend that the CSST NIR imager takes at least 30 frames in the non-destructive readout mode during its nominal 150~s exposures.

Compared to the last frame, the quasi-optimal stacking method requires a lower count of photoelectrons to achieve the same SNR, which is equivalent to a reduction of the readout noise. To be quantitative, we define an effective readout noise as follows:
\begin{equation}\label{eq_snrnobj}
    r_\mathrm{eff} = \sqrt{\frac{1}{n_\mathrm{pix}}\left[\left(\frac{s_\mathrm{80}}
    {\mathrm{SNR}}\right)^2-s_\mathrm{80}\right]-b}, 
\end{equation}
$n_\mathrm{pix}$ is the pixel number within $R_{80}$. 
The results are also shown in \TabRef{tab3}. Stacking $30$ frames results in the effective readout noise decreased by about $62\%$, while stacking $70$ frames further decreases it by about $71\%$. 
Note that these performance improvements are achieved by data reduction with no extra hardware work.

\section{Conclusion}
\label{sect5}

In this work, we develop a quasi-optimal stacking method for ramp images taken in non-destructive readout mode. It is designed to enhance faint object detection by setting a low-SNR target case for optimization. 
Although the stacking weights obtained are truly optimal only for objects of uniform brightness at a specific $\mathrm{SNR_{target}}$ per pixel in the last frame, the stacking results of flat-field ramp images and point-source ramp images are not very sensitive to $\mathrm{SNR_{target}}$ as long as it is kept low. 
A good choice is $\mathrm{SNR_{target}}=1$. 
The quasi-optimal stacking method outperforms the other methods tested in \SecRef{sect3} except that, in the photon-noise dominant regime, it is eventually surpassed by the last frame at a sufficiently high SNR. 
Because of its ability to enhance the SNR in the low-SNR regime, the quasi-optimal stacking method can recoup a sizeable sample of faint objects of interest that are otherwise undetectable and help improve our knowledge about the faint-end luminosity function of such objects.

 Using the CSST NIR imager as an example, we estimate that the quasi-optimal stacking method can improve its limiting magnitudes in both $\mathrm{J'}$ and $\mathrm{H'}$ bands by about 0.5~mag (0.6~mag) and decrease its effective readout noise by about 62\% (71\%) with 30 (70) frames taken in the non-destructive readout mode within the nominal exposure time of 150~s. 
 These are highly significant improvements, and the reduction method is worth implementing in the CSST NIR data pipeline in the future.

\section*{acknowledgements}
This work was supported by the National Key R\&D Program of China No. 2022YFF0503400, the National Natural Science Foundation of China grant U1931208 and China Manned Space Program through its Space Application System. 

\section*{Author contributions}
Hu Zhan conceived the ideas and designed the study. Guanghuan Wang implemented the study.
Guanghuan Wang and Hu Zhan wrote the paper with equal contribution. 
Zun Luo, Chengqi Liu and Youhua Xu assisted on simulations and data processing and made suggestions on the writing. 
Chun Lin, Yanfeng Wei and Wenlong Fan provided detailed characteristics of NIR detectors and provided useful comments on the manuscript. 
All authors read and approved the final manuscript. 

\section*{Declaration of Interests}
The authors declare no competing interests.
Hu Zhan was not involved in the review process of this paper. 

\bibliographystyle{ati} 
\bibliography{ati}
\end{document}